\documentclass[%
superscriptaddress,
showpacs,
amsmath,amssymb,
aps,
prc,
showkeys,
twocolumn,
floatfix
]{revtex4-1}
\usepackage{amssymb}

\usepackage{amsmath,mathrsfs,amssymb}
\usepackage{eurosym}
\usepackage{graphicx,subfigure}
\usepackage{color}
\usepackage{indentfirst}
\usepackage{extarrows}
\usepackage{hyperref}
\hypersetup{colorlinks=true, citecolor=blue, urlcolor=blue, linkcolor=blue}
\usepackage{ulem}

\begin{document}

\title{Status of deep subbarrier $\mathbf{{}^{12}{\rm C}+{}^{12}{\rm C}}$ fusion and advancing the Trojan horse method}

\author{A. M.  Mukhamedzanov}
\address{Cyclotron Institute, Texas A$\&$M University, College Station, Texas, 77843, USA \\
E-mail: akram@comp.tamu.edu}

\begin{abstract}
In this paper,  I will update the current status of the carbon-carbon fusion research taking into account that after the latest analysis 
[Beck {\it et al.}  Eur. Phys. J. A  {\bf 56}, 97  (2020), Letter to the Editor]  new important experimental and theoretical results had been published and will discuss how to advance new THM measurements to extract  the low-energy astrophysical $S$-factors.
\end{abstract}

\maketitle

\section{Introduction}

Carbon plays a key role in astrophysics, and information about nuclear reactions involving carbon is essential 
for understanding nucleosynthesis in stars. Moreover, two nuclear astrophysical processes involving carbon, 
${}^{12}{C}(\alpha,\gamma){}^{16}{\rm O}$ and ${}^{12}{\rm C}+{}^{12}{\rm C}$ fusion remain one of the main focus 
of scientists for many years. While our knowledge about the first reaction has been significantly improved, 
the implication of the carbon-carbon fusion reaction on stellar evolution is still under debate
despite extensive research, both theoretical and experimental, see  \cite{Mori,Becker,Kettner,Aguilera,Spillane,Tumino,Jiang,Patterson,Esbensen,Wiescher,Bonasera,Godbey,Wen,SPP,Chen,Khoa,Notani,Jiang13,Assuncao,Zhang2019,Nagatani}.

The ${}^{12}{\rm C} + {}^{12}{\rm C}$ fusion reaction has different exit channels but five exothermic channels attract the attention of the scientists. These are
\begin{widetext}
\begin{align}
{}^{12}{\rm C} + {}^{12}{\rm C} \to {}^{23}{\rm Na}(3/2^{+},\,E_{x}=0.0\, {\rm MeV}) + p_{0}, \quad  Q= 2.241\, {\rm MeV},   \\
{}^{12}{\rm C} + {}^{12}{\rm C} \to {}^{23}{\rm Na}(5/2^{+},\,E_{x}=0.44\,{\rm MeV}) + p_{1}, \quad Q= 1.801\, {\rm MeV},    \\
{}^{12}{\rm C} + {}^{12}{\rm C} \to {}^{20}{\rm Ne}(0^{+},\,E_{x}=  0.0 \,{\rm MeV}) + \alpha_{0}, \quad Q= 4.617\, {\rm MeV}, \\  
{}^{12}{\rm C} + {}^{12}{\rm C} \to {}^{20}{\rm Ne}(3/2^{+},\,E_{x}=1.634\, {\rm MeV}) + p_{0}, \quad Q= 2.983\, {\rm MeV},\\
 {}^{12}{\rm C} + {}^{12}{\rm C} \to {}^{24}{\rm Mg}(0^{+},\,E_{x}=0.0\, {\rm MeV}) + \gamma, \quad  Q= 13.933\, {\rm MeV}.
\label{12C12Creactionchannels1}
\end{align}
\end{widetext}
For stars with masses $(8-10)m_{\odot}$ the effective energy interval (Gamow window)  for the carbon-carbon fusion is $ E \sim 1.2-1.8$ MeV, where 
$E \equiv E_{{}^{12}{\rm C}\,{}^{12}{\rm C}}$ is the relative ${}^{12}{\rm C}-{}^{12}{\rm C}$ kinetic energy. It corresponds to the temperature interval $(0.35-0.64 )T_{9}$  (temperature $T_{9}= 10^{9}$ Kelvin). 
 For heavier stars with masses $\sim 25\,m_{\odot}$ the effective 
energy interval is $E \sim 1.8-2.6$ MeV corresponding to $(0.64-1.15)T_{9} $. 

The Coulomb barrier for the carbon-carbon head-on  collision (calculated for the carbon radius $1.4\,\times12^{1/3}$ fm 
neglecting the nuclear interaction)  is $\approx 6.6$ MeV making the astrophysically relevant energies deep sub-Coulomb. The lowest measured resonance energy $E \approx 2.14$ MeV was reported in \cite{Spillane}. However, large uncertainties made it impossible to determine the strength of this resonance. 
Although it is very probable that the accuracy of direct measurements in the region $E \sim 2.0$ MeV soon will be improved, 
it is not feasible, at least in the near future, that direct measurements can cover the whole astrophysically relevant for masses $(8-10)m_{\odot}$ energy interval  $1.2-1.8$ MeV.  

The first experimental breakthrough was presented in \cite{Tumino}. In this work, the indirect Trojan horse method (THM) utilizing the reaction generated by the collision of ${}^{14}{\rm N}$ with ${}^{13}{\rm C}$ was used to obtain 
the astrophysical $S$-factor for the carbon-carbon fusion down to $E=0.8$ MeV. The measured astrophysical factor showed pronounced low-energy resonances at $E=0.88,\,0.98$ and $1.5$ MeV accompanied by quite a few more minor resonances
demonstrating the power of the THM. The advantage of the THM is the absence of the Coulomb penetrability factor in the $\,{}^{12}{\rm C}+{}^{12}{\rm C}\,$ channel because the nucleus $\,{}^{12}{\rm C}\,$ emerging from the projectile $\,{}^{14}{\rm N}\,$ is the off-the-energy shell (off-shell). 
It allows one to extract the low-energy $S$-factor avoiding the challenging problems appearing in direct measurements. However, the analysis of the THM reactions is complicated and is not as straightforward as for direct measurements. 
The sharp rise of the astrophysical factor toward low energies obtained in \cite{Tumino}
was the result of neglecting the Coulomb interaction in the initial and, especially, the intermediate and final states of the THM reactions \cite{muk2019,muk2020, Beck}, see also discussion below. Taking into account the Coulomb-nuclear distortions decreases the extracted in \cite{Tumino} astrophysical factors, including the resonance peaks by up to three orders of magnitude.  

From the theoretical point of view, existing of the resonances in the astrophysical factors of the ${}^{12}{\rm C} + {}^{12}{\rm C}$ fusion is one of the most exciting topics in nuclear theory related with nuclear astrophysics. These resonances are believed to be molecular ${}^{12}{\rm C}+{}^{12}{\rm C}$ configurations in the compound ${}^{24}{\rm Mg}$. 
Given the importance of the issue, it is not surprising that there have been so many theoretical papers devoted to carbon-carbon fusion. Below I will briefly discuss some most advanced publications related to carbon-carbon fusion.

In \cite{Jiang},   different methods of extrapolation  of astrophysical factors down to astrophysically relevant energies   for $\,{}^{12}{\rm C}\,$ and $\,{}^{16}{\rm O}\,$  fusion reactions  were exploited.  The authors came to the conclusion that the fusion hindrance (decrease of the $S$-factor toward low energies)  should be taken into account.  However, the physical  nature of the hindrance effect was not clarified.

 The paper \cite{Esbensen} presents one of the most advanced studies of carbon isotopes fusion. The fusion cross-sections were calculated by solving the coupled-channel equations using the Woods-Saxon or M3Y potentials augmented with a repulsive core allowing one to take into account nuclear incompressibility. 
The mutual excitations of both colliding nuclei were taken into account. The nuclear excitations of both colliding carbon nuclei were obtained from the surface excitations, which is described by the radial deformation depending on deformation amplitudes. Two different results corresponding to ten and twelve 
coupled channels were presented. The fusion cross-sections demonstrated 
smooth behavior and exceeded experimental data, especially for twelve channels. They can be considered as an upper limit of the experimental cross-sections.  These calculations were not intended to reproduce resonance structure because the cluster effects were not included in the formalism.
 
In \cite{Wiescher}, the quantum dynamical model based on the solving of the time-dependent Schr\"odinger 
equation with a collective Hamiltonian. The model used the Time-Dependent Wave-Packet method. The calculated $S$-factor, in contrast to the potential and coupled-channel methods, shows three resonances at $E > 4$ MeV but smooth behavior at lower energies. The absence of resonance structures at low energies was due to the lack of cluster effects. 

In \cite{Bonasera} the carbon-carbon fusion cross-section was also calculated using the TDHF method based on the Feynman Path Integral Method. 
The general trend of the modified astrophysical factor showed a smooth decrease toward low energies. No resonance structure was obtained because the model did not include cluster effects.
To reproduce the low-energy resonances, the authors engaged the Bass ion-ion potential, which contains nuclear, Coulomb, and centrifugal terms. Its strength was adjusted to produce the resonances 
found in the THM experiment \cite{Tumino}. These resonance can be considered as phenomenological fit of the experimental data.
 It should be underscored that the potential model was utterly independent of the basic TDHF approach, and the obtained resonances were added to the TDHF calculations.

 To analyze the ${}^{12}{\rm C} + {}^{12}{\rm C}$ fusion two microscopic methods based on density-constrained static Hartree-Fock and density-constrained TDHF  methods were applied in \cite{Godbey}. 
The bare ion-ion potentials were derived using the density energy functional for each method. 
The calculated $S$-factors using different microscopic approaches showed similar behavior with an apparent rise toward low energies contradicting the hindrance model \cite{Jiang}. It was underscored that the $\alpha$-clustering effects in ${}^{12}{\rm C}$, which were not taken into account, can influence the fusion process. 

In Ref. \cite{Wen}, published two years later, the hindrance effect was observed by solving the coupled-channels approach with improved incoming boundary conditions. The ion-ion potential included Coulomb and Woods-Saxon nuclear potentials. The paper presented some insight into the hindrance phenomenon.
Again, a smooth energy behavior of the modified $S$-factor without any resonance structure  with a drop toward low energies was found. 

The origin of the resonance structure of the ${}^{12}{\rm C}+ {}^{12}{\rm C}$ fusion $S$-factor was analyzed in \cite{Jiang13}. The fusion of these two carbon nuclei is compared with fusion of ${}^{12}{\rm C}+ {}^{13}{\rm C}$ 
and ${}^{13}{\rm C}+ {}^{13}{\rm C}$ systems. The conclusion of the paper was that nonoverlapping compound 
states of ${}^{24}{\rm Mg}$ at energies $E < 7$ MeV (the scantiness of the resonances in the ${}^{12}{\rm C}+ {}^{12}{\rm C}$ system and the narrow resonance widths make them nonoverlapping) caused the resonance structure and suppression of the deep sub-barrier fusion of ${}^{12}{\rm C}+ {}^{12}{\rm C}$. 
The calculated $S$-factors for the Woods-Saxon and the coupled channels method using the M3Y+repulsion potential did not show any resonance structure. 

Another coupled-channel method was exploited in \cite{Assuncao}. The ${}^{12}{\rm C}+ {}^{12}{\rm C}$ fusion 
modified $S$-factors were calculated using the multichannel folding model. Up to four excited states of ${}^{12}{\rm C}$, $(0_{1}^{+},\,2^{+},\,0_{2}^{+},\,3^{-})$, had been taken into account. The folding potential was constructed using the $M3Y$ $\,NN\,$ potential and ${}^{12}{\rm C}$ densities using the resonating group method. For the number of the excited states $\leq 2$ the calculated modified $S$-factors showed monotonic increase as energy decreases. 
No resonance structure appeared because the approach included no cluster effects.

To summarise, the key point is that all the above-cited works did not include the cluster degrees of freedom in ${}^{12}{\rm C}$. The resulting astrophysical factors demonstrated a smooth energy behavior, which depends on the adopted model. The trends toward low energies varied from increase to hindrance. The dependence of the modified $S$-factors on the number of the included excited states was demonstrated.  I mean that
 neither of the referred models was capable of reproducing the resonance structure of the ${}^{12}{\rm C}+ {}^{12}{\rm C} 
\to \alpha + {}^{20}{\rm Ne}$ and ${}^{12}{\rm C}+{}^{12}{\rm C} \to p +{}^{23}{\rm Na}$ reactions, because these reactions are many-nucleon rearrangement ones proceeding through the intermediate resonance ${}^{24}{\rm Mg}^{*}$.
The study of resonances in these reactions requires a microscopic approach, which considers nucleon degrees of freedom and channel coupling. 

In 2021, the first fully microscopic approach based on the antisymmetrized molecular dynamic (AMD) had been published \cite{TaniguchiKimura}. It was a real breakthrough in the long-stalled theory of carbon-carbon fusion. 
The authors presented the first fully microscopic treatment of the ${}^{12}{\rm C}+{}^{12}{\rm C}$ fusion without any adjustable parameters. The AMD nuclear wave function is given by the parity-projected Slater determinant,
whose elements are nucleon Gaussian wave packets. The superposition of the wave functions of different rearrangement channels formed as the result of the carbon-carbon fusion 
 described the resonances of the compound nucleus ${}^{24}{\rm Mg}$. The resonances in the $S$-factor appear only due to the channel coupling and disappear if only one channel ${}^{12}{\rm C}+{}^{12}{\rm C}$
is taken into account.

In this paper,  the current status of the carbon-carbon fusion is updated, taking into account that after the latest analysis 
\cite{Beck} new important experimental and theoretical results had been published. I will discuss how to advance new THM measurements to extract the low-energy astrophysical $S$-factors discussing the kinematics of two THM reactions:  ${}^{12}{\rm C}({}^{14}{\rm N},\,d){}^{24}{\rm Mg}^{*}$ and  ${}^{12}{\rm C}({}^{13}{\rm C},\,n){}^{24}{\rm Mg}^{*}$.

In what follows the system of units in which $\hbar=c=1$ is used. 

\section{Latest experimental and theoretical results}

\begin{figure}[b]
\includegraphics[width=3.7 in,height=6.0in]{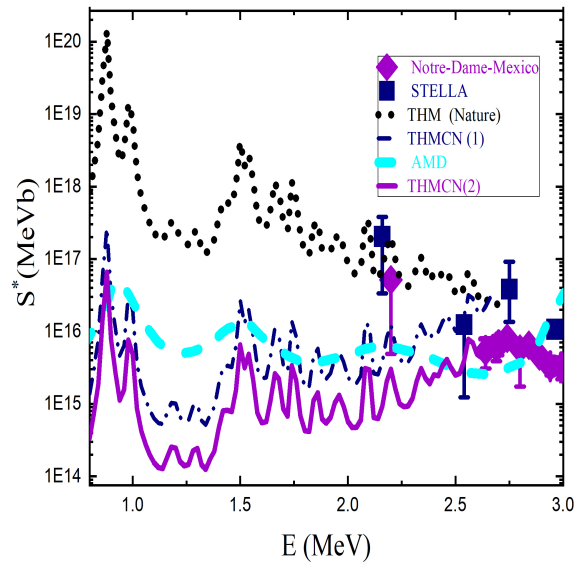}
  \caption{ Modified astrophysical factors for the ${}^{12}{\rm C} + {}^{12}{\rm C}$ fusion. The   navy rectangles and the magenta diamonds are the astrophysical factors from \cite{STELLA} (STELLA) and \cite{Tan}  (Notre-Dame-Mexico), black dotted line is the THM $S^{*}(E)$-factor \cite{Tumino}   (THM (Nature)),  the navy  dash-dotted line is the Coulomb-nuclear renormalized astrophysical factor \cite{muk2019,muk2020} normalized to the $S$-factor from 
  \cite{STELLA}  at $E=2.54$ MeV  (THMCN(1)), the magenta solid line is the Coulomb-nuclear renormalized astrophysical factor normalized to the $S$-factor from \cite{Tan}  at $E=2.64$ MeV  (THMCN(2)),  the cyan dash  line is the AMD microscopic calculations from \cite{TaniguchiKimura}.}
\label{fig_Sctrs1}
\end{figure}

Fig. \ref{fig_Sctrs1} shows the modified astrophysical  factors for the carbon-carbon fusion published during the last three years.
First of all, in 2020, two important direct measurements of the carbon-carbon fusion  were published  by STELLA collaboration \cite{STELLA} and Notre Dame-Mexico \cite{Tan}. These experiments show some disagreement between both measurements, see Fig. \ref{fig_Sctrs1}. The lowest measured energies in both experiments are $E \sim 2.2$ MeV. Thus, since 2007, when the energy of $E=2.14$ MeV was reached in \cite{Spillane}, direct measurements were not able to measure the $S$-factors at energies below  2.1 MeV.
Hence it is unlikely that soon direct measurements will reach energies down to $E \sim 1.5$ MeV.  

Also in Fig. \ref{fig_Sctrs1}  is presented  the first  modified astrophysical $S^{*}(E)$-factor obtained  using the indirect THM measurements \cite{Tumino}, which reached the lowest energy $\,E=0.8 $ MeV. 
Since the width of the Gamow window for the carbon-carbon 
is comparable or wider than the resonance widths, the averaged modified astrophysical factor 
\begin{align}
S^{*}(E)= \sigma(E)\,E\,e^{(\frac{87.21}{\sqrt{E}}  + 0.46\,E)},
\label{Sgen1}
\end{align}
where $\sigma(E)$ is the fusion cross-section, is a relevant approach for the astrophysical application. 
As one can see, the original THM data show three significant resonance peaks with a sharp increase toward small energies. However, as we have discussed, the analysis of the THM data was performed in the plane-wave approximation (PWA), neglecting the Coulomb rescattering of the nuclei in the initial and, especially, in the intermediate and final channels. 
 At the normalization energy of the THM data to the direct ones $E=2.67\,$ MeV the relative $\,d-{}^{24}{\rm Mg}\,$ kinetic energy is $0.92$ MeV with the Coulomb barrier about $3$ MeV. Moreover, although the relative ${}^{14}{\rm N}- {}^{12}{\rm C}$ kinetic energy $13.84$ MeV in the entry channel of the THM reaction is higher than the Coulomb barrier of $10$ MeV, the Coulomb parameter in the entry channel is $4.4$. Such a strong Coulomb interaction in the initial channel and the deep sub-Coulomb energy in the intermediate channel explains the failure of the  PWA used in \cite{Tumino}. Including the Coulomb-nuclear distorted waves causes a drop of the low-energy astrophysical factor by about of factor $10^{3}$, see Fig. \ref{fig_Sctrs1}. 

In Fig. \ref{fig_Sctrs1} are also presented the two THM modified astrophysical factors $S^{*}(E)$-factors obtained from data in \cite{Tumino} including the Coulomb-nuclear distortions. These distortions are taken into account by replacing the PWA transfer differential cross-sections used in \cite{Tumino} with the DWBA one. 
 The dash-dotted navy curve is normalized to the $S^{*}$-factor from \cite{STELLA}  at $E=2.54$ MeV (THMCN(1)), while the solid magenta line is the same Coulomb-renormalized THM astrophysical factor but normalized to the astrophysical factor from \cite{Tan} at $E=2.64$ MeV  (THMCN(2)).  

The calculated AMD $S^{*}$-factor \cite{TaniguchiKimura}, cyan dash line,  shows strong resonances at $0.94$ MeV and $1.5$ MeV and a resonance at $2.2$ MeV, see the cyan dash curve in Fig.\ref{fig_Sctrs1}. Moreover, the AMD astrophysical factor qualitatively agrees with the THMCN $S^{*}$-factors, especially with THMCN(1). 
Further improvement of the indirect THM measurements and updated AMD approach can converge.

\section{Advancing indirect THM}

\subsection{Energy-momentum equations for THM}
\label{Energyequations1}

The first indirect THM measurements \cite{Tumino} demonstrated that using this technique scientists can measure the astrophysical factor for ${}^{12}{\rm C}+ {}^{12}{\rm C}$ down to $E=0.8$ MeV covering the whole energy interval relevant for nuclear astrophysics. 
The THM experiment's positive outcome is discovering the resonances at energies $E < 2$ MeV. 
However, the first indirect experiment and its analysis encountered a few weighty drawbacks \cite{muk2019,muk2020, Beck}. It resulted in a wrong energy track of the THM astrophysical factor and resonance spins assignment. 
In what follows, I will discuss how to advance the indirect THM to finally solve the problem of the low-energy astrophysical factor for the carbon-carbon fusion. 
To help the reader, I present some essential equations needed for the THM application. 
Let us consider the THM reaction
\begin{align}
a+ A \to s+F^{*} \to s+ b + B,
\label{THMreaction1}
\end{align}
where $a=(s\,x)$ is the Trojan horse (TH) particle, which brings the particle $x$, hidden inside,  behind the $a+A$ barrier allowing it to interact with the target $A$ while particle  $s$ leaves as a spectator,  and $F^{*}$ is the resonance in the subsystem $F=(xA)$.  
The idea of the THM is to extract the information about the binary resonant subreaction 
\begin{align}
x+ A \to b+B.
\end{align}

\begin{figure}
\includegraphics[width=3.0in,height=3.5in]{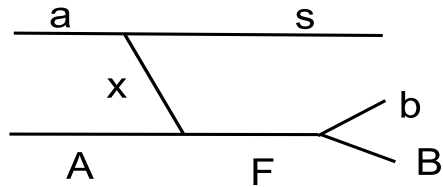}
  \caption{The mechanism of the THM in the PWA.}
\label{fig_PWAdiagram}
\end{figure}

The THM reaction is a two-step reaction proceeding through the intermediate resonance. Its mechanism is described by the diagram depicted in Fig. \ref{fig_PWAdiagram}.
Note that when calculating  the THM reaction amplitude, a priori, the distorted waves in the initial, intermediate and final states should be taken into account \cite{mukkad2019}. 

The first step is the transfer reaction $a+A \to s+F^{*}$
populating the resonance state $F^{*}$, which on the second stage decays into the two-body channel $b+B$. 
In what follows, I first recall some basic equations and definitions  which are used in the THM analysis.

The energy conservation in the center-of-mass (c. m.) of the TH reaction reads
\begin{align}
E_{aA} - \varepsilon_{sx} = E_{sF} + E= E_{sF} + E_{bB} - Q_{if},
\label{enconservTH1}
\end{align}
where $E \equiv E_{xA},\,$ $\,E_{j\,t}= k_{j\,t}^{2}/(2\,\mu_{j\,t}),\,$  $\,Q_{if}= m_{x}+ m_{A} - m_{b} - m_{B}\,$, $\,E_{j\,t}$, $\,{\rm {\bf k}}_{j\,t}\,$ and $\,\mu_{j\, t}\,$ are the relative kinetic energy, on-the-energy shell relative momentum and reduced mass of the particles $j$ and $t$, $\,m_{j}$ is the mass of the particle $j$, $\,\varepsilon_{sx}\,$ is the binding energy of the particles $s$ and $x$ in the TH particle $\,a=(sx)$.  
$\;E_{R(xA)} =E_{0(xA)} - i\,\Gamma/2,\,$  is a resonance energy in the subsystem $x+A$,  $\,E_{0(xA)}\,$ is the real part of the resonance energy in the channel $x+A$, $\,\Gamma\,$ is the total resonance width of the resonance $F^{*}$ populated in the transfer reaction.

We consider a two-state coupled channel problem in which the resonance formed in the channel $i=x+A$ decays into a different channel $f=b+B$. Therefore, when in the channel $i$ $\;E \to E_{R(xA)}\,$ the relative energy $E_{bB}$ approaches the resonance energy $\,E_{R(bB)}\,$ in the channel $f$:  $\;E_{R(bB)}= E_{0(bB)} - i\,\Gamma/2$.

For $E \to E_{R(xA)}$, due to energy conservation, see (\ref{enconservTH1}), one gets that $E_{sF} \to E_{R}$, where
\begin{align}
E_{R}= E_{0(sF)} - i\,\Gamma/2.
\label{ER1}
\end{align}
Here
\begin{align}
E_{0(sF)}= E_{aA} - \varepsilon_{sx} - E_{0(xA)} = E_{aA} -\varepsilon_{sx} + Q_{if} - E_{0(bB)}
\label{E01}
\end{align}
is the real part of the resonance energy in the system $s+F$.

From the energy-momentum conservation in the three-ray vertices $a \to s+x$ and $x+A \to F^{*}$ in the diagram of Fig. \ref{fig_PWAdiagram} follows that  $xA$ relative kinetic energy $E$ is
\begin{align}
E= \frac{p_{xA}^{2}}{2\,\mu_{xA}} -   \frac{p_{sx}^{2}}{2\,\mu_{sx}} - \varepsilon_{sx}.
\label{ExApxApsx1}
\end{align}
In the system $k_{a}=0$  Eq. (\ref{ExApxApsx1})  reduces to
\begin{align}
E= \frac{m_{x}}{m_{xA}}\,E_{A} - \frac{k_{s}^{2}}{2\,\mu_{sF}} +  \frac{{\rm {\bf k}}_{s} \cdot  {\rm {\bf k}}_{A} }{m_{xA}} - \varepsilon_{sx}.
\label{ExAEA1}
\end{align}
 ${\rm {\bf p}}_{j\,t}$  is the $j-t$ the Galilean-invariant relative momentum when one of the particles $j$ or $t$ is virtual (off-the-energy shell). 
\begin{align}
{\rm {\bf p}}_{xA}=  -{\rm {\bf k}}_{A} +  \frac{m_{A}}{m_{xA}}\,{\rm {\bf k}}_{F} ,  \nonumber\\
{\rm {\bf p}}_{sx}=   {\rm {\bf k}}_{s} - \frac{m_{s}}{m_{sx}}\,{\rm {\bf k}}_{a}.             
\label{pxApsx1}
\end{align}
$\,m_{j\,t}= m_{j}+m_{t}$, $\,\mu_{j\,t}= m_{j}\,m_{t}/m_{j\,t},\,$  ${\rm {\bf k}}_{j}$ is on-the-energy shell momentum of particle $j$. 
In the  c. m. of the THM reaction ${\rm {\bf k}}_{a}= - {\rm {\bf k}}_{A}$
and ${\rm {\bf k}}_{s}= - {\rm {\bf k}}_{F}$ and ${\rm {\bf p}}_{xA}=  {\rm {\bf k}}_{a} -  \frac{m_{A}}{m_{xA}}\,{\rm {\bf k}}_{s}$.

For on-the-energy shell particles 
\begin{align}
{\rm {\bf k}}_{j\,t}=  \frac{m_{t}\,{\rm{\bf k}}_{j}-  m_{j}\,{\rm {\bf k}}_{t}}{m_{j\,t}}
\label{kij1}
\end{align}
and ${\rm {\bf k}}_{aA}= {\rm {\bf k}}_{a}, \;\; {\rm{\bf k}}_{sF}={\rm {\bf k}}_{s}$.

We can express $p_{sx}$ and $p_{xA}$ in terms of $E$. In the c. m. of the THM reaction  we get from Eq. (\ref{pxApsx1}) 
\begin{widetext}
\begin{align}
&p_{sx}= \sqrt{  2\,\mu_{sF}\,(E_{aA} - E  - \varepsilon_{sx} )  - 2\,\frac{m_{s}}{m_{a}}\,\cos(\theta)\, \sqrt{2\,\mu_{sF}\,(E_{aA} - E  - \varepsilon_{sx} )}\,\sqrt{2\,\mu_{aA}\,E_{aA}}  + \big( \frac{m_{s}}{m_{a}}\big)^{2}\,2\,\mu_{aA}\,E_{aA} },  \nonumber\\
&p_{xA}= \sqrt{  2\,\mu_{aA}\,E_{aA} - 2\,\frac{m_{A}}{m_{xA}}\,\cos(\theta)\, \sqrt{2\,\mu_{sF}\,(E_{aA} - E  - \varepsilon_{sx} )}\,\sqrt{2\,\mu_{aA}\,E_{aA}}  + \big( \frac{m_{A}}{m_{xA}}\big)^{2}\,2\,\mu_{sF}\, (E_{aA} - E  - \varepsilon_{sx} )  }.
\label{psxpxAEEaA1}
\end{align}
\end{widetext}
where $\cos(\theta)= {\rm {\bf {\hat k}}}_{s} \cdot {\rm {\bf {\hat k}}}_{a}$, $\,{\rm {\bf {\hat k}}}= {\rm {\bf k}} /k$.

Equation  (\ref{ExAEA1})  shows that  at fixed  $\,{\rm {\bf k}}_{A}\,$  to  vary  $\,E\,$  one needs to vary ${\rm {\bf k}}_{s}$.

\subsection{ THM triple differential cross section}
\label{TripleDCS1}

Let me consider now  another important equations, such as differential cross sections (DCSs), triple, double and single, and THM astrophysical factor. 
I start from the general equation for the triple DCS  for the reaction $2\,\, {\rm particles} \to 3\,\, {\rm particles}$.
The kinematics of these reactions is determined by the four momentum-energy conservation equations:
\begin{align}
&{\rm {\bf k}}_{a} + {\rm {\bf k}}_{A}= {\rm {\bf k}}_{s} + {\rm {\bf k}}_{b}+ {\rm {\bf k}}_{B},  \nonumber\\
&E_{a}+E_{A} + m_{a}+m_{A}= E_{s}+E_{b}+E_{B} + m_{s}+ m_{b}+ m_{B}.
\label{Energymomentum1}
\end{align}
The total number of the variables describing three particles in the final state is nine. Subtracting four variables from Eqs. (\ref{Energymomentum1}) we get five independent variables in the THM reactions. The incident beam selects a direction (axis) in the space. The angle describing the rotation around this axis can be dropped. Then there are only four significant independent Galilean-invariant variables, which are needed to describe the THM triple differential cross-section. These four variables, for example, can be the energies and scattering angles of two particles. They provide the full 3-body kinematics. 

Here  we use another set of independent variables using the fact  that the THM reaction is the two-step process. 
As independent variables we select  $E_{sF},\,$ solid angles  $\,\Omega_{{\rm {\bf k}}_{sF}},\,$ and $\,\Omega_{{\rm {\bf k}}_{bB}}.\,$  Then the triple DCS  is given by \cite{mukkad2019,Dolinski}
\begin{align}
\frac{{{{\rm d}^3}\sigma }}{{{\rm d}{\Omega _{{k_{bB}}}}{\rm d}{\Omega _{{k_{sF}}}}{\rm d}{E_{sF}}}} = |N_{C}|^{2}\,\frac{\mu _{aA}\,\mu _{sF}}{{{{(2\pi )}^3}}}\,\frac{ k_{sF}}{k_{aA}}\,\frac{ k_{bB} }{\mu_{bB}}\,{\overline {\big|{\cal M}_{ R}\big|^{2}}},
 \label{TripleDCSspin1}
\end{align}
where $k_{sF} = \sqrt{2\,\mu_{sF}\,E_{sF} }$.

\subsubsection{Multi-level, two-channel case}
\label{multilevel1}

To write done explicitly  ${\overline {\big|{\cal M}_{ R}\big|^{2}}}$,  first,  I consider a multilevel, two-channel  case.
This case should be considered when there are overlapping resonances with the same quantum numbers.  Such resonances interfere and the best way to analyze such cases is to use the $R$-matrix method: 
\begin{align}
&\overline {\big|{\cal M}_R{\big|^2}} =\,{ {\widehat J}_{a} {\widehat J}_{A}  }\sum\limits_{ {M_B}{M_b}{M_s}{M_a}{M_A} }\,\Big| \sum\limits_{M_{F_{\nu}}\,M_{F_{\tau}  }  } \sum\limits_{\nu\,\tau=1}^{N}\,
 \nonumber\\
& \times W_{\nu\,M_{B} M_{b} }^{M_{F_{\nu}}} ({\rm {\bf k}}_{bB})\,  [{\bf A}]_{\nu\,\tau}^{-1}\,M_{\tau\, M_{F_{\tau}} M_{s}; M_{A}M_{a}}({\rm {\bf {k}}}_{sF},\,{\rm {\bf k}}_{aA})\Big|^{2}.
 \label{multilevelMR1}
 \end{align}
 $\,J_{i}(M_{i})$ is the spin  (its projection) of nucleus $i$,  $\,{\hat J}_{i}= 2\,J_{i}+1.\,$
$N$ is the number of $R$-matrix levels and ${\bf A}$ is the $R$-matrix level matrix, which provides the coupling of the different levels and channels:
\begin{align}
&{\bf A}_{\nu\,\tau}(E)=  \big(E_{\nu}  - E \big)\,\delta_{\nu\,\tau} - \sum\limits_{c}\,\gamma_{\nu\,c}\,\gamma_{\tau\,c}\Big[{\hat S}_{c}(E) -B_{c}          \nonumber\\
& +i\,P_{c}(E,\,R_{ch}) \Big].
\label{Levelmatrix A1}
\end{align}
Here $ \sum\limits_{c}$ is the sum over the included open channels. $\,\gamma_{\tau\,c}\,$ is the formal $R$-matrix  reduced width amplitude of the level $\tau$ in the channel $c$, 
${\hat S}_{c}(E),\,$ $B_{c}$ and $P_{c}(E,\,R_{ch})$ are the $R$-matrix level shift,  boundary condition and penetrability factor in the channel $c,\,$ respectively.  $E_{\tau}$ is the $\tau$-th level energy.  

The formal $R$-matrix reduced  width amplitude  $\,\gamma_{\tau\,c}$, which is a fitting parameter, is related to the formal $R$-matrix resonance width $\Gamma_{\tau\,c}(E)$:
\begin{align}
\Gamma_{\tau\,c}(E) = 2\,P_{c}(E,R_{ch})\,\gamma_{\tau\,c}^{2},
\label{GammaxAgamma1}
\end{align}
\begin{align}
P_{l_{c}}(E,\,R_{ch}) = k_{c}\,R_{ch}\,|O_{l_{c}}(k_{c}\,R_{ch})|^{-2},
\label{PlxA1}
\end{align}
where $l_{c}$ and $\,k_{c}$ are the relative orbital angular momentum and the relative momentum in the channel $c$.
I need to add a few words about the $R$-matrix energy levels $E_{\tau}$. One can adopt one of the energy levels equal to known resonance energy, and then all other energy levels will fitting parameters

Note that the observable ${\tilde \Gamma}_{\tau\,c}(E_{0(c)}) $ and the formal $\Gamma_{\tau\,c}(E_{0(c)})$ partial resonance widths at the real part of the resonance energy $E_{0(c)}$ in the channel $c$  are related by
\begin{align}
{\tilde \Gamma}_{\tau\,c}(E_{0(c)}) =  \frac{\Gamma_{\tau\,c}(E_{0(c)}) }{1+ \sum\limits_{c}\,\gamma_{\tau\,c}^{2}\,\frac{{\rm d}{\hat S}_{c}}{{\rm d}E}|_{E=E_{0(c)}}  }.
\label{ObsGamma1}
\end{align}

$M_{\tau\, M_{F_{\tau}} M_{s}; M_{A}M_{a}}({\rm {\bf k}}_{sF},{\rm {\bf k}}_{aA})$  is the amplitude of the transfer reaction $a+ A \to s+ F_{\tau}^{*}$ populating the resonance state $F_{\tau}^{*}$:
\begin{widetext}
\begin{align}
&M_{\tau\, M_{F_{\tau}}\,{M_s}; M_A\, M_a}({\rm {\bf k}}_{sF},\,{\rm {\bf k}}_{aA}) =  {i^{ - {l_{xA}}}} {e^{ - i\,{\delta_{l_{xA}} ^p}(k_{xA})}}\,\frac{1}{2\,\mu_{xA}}\,\sqrt{\frac{\Gamma _{\tau (xA)}\,\mu _{xA} }{\,k_{xA}}}\,O_{{l_{xA}}}({k_{xA}}{R_{ch}})\,j_{{l_{xA}}}({k_{xA}}{R_{ch}})               \nonumber\\
& \times {\cal W}_{l_{xA}}\,{Y_{{l_{xA}},{m_{{l_{xA}}}}}}({\widehat {\bf{k}}_{xA}})\,M_{{\tau\,M_{F_{\tau}}}{M_s};{M_A}{M_a}}^{DWZR(prior)}({\rm {\bf  k}}_{sF},\,{\rm {\bf k}}_{aA}),
\label{MDWBACoul2}
\end{align}
\end{widetext}
where the zero-range prior DWBA amplitude is
\begin{align}
& M_{\tau\,M_{F_{\tau}}{M_s};{M_A}{M_a}}^{DWZR(prior)}({{\bf{k}}_{sF}},{{\bf{k}}_{aA}}) =\sum\limits_{ m_{s_{xA}}\,m_{l_{xA}}M_{x} }\,
C_{ s_{xA}m_{s_{xA}}\,\,l_{xA}m_{l_{xA}}}^{ J_{F}M_{F_{\tau}}}                                                            \nonumber\\
&\times C_{J_{x}M_{x}\,\,J_{A}M_{A} }^{s_{xA}m_{s_{xA}}}                      
C_{s_{sx}m_{s_{sx}}\,l_{sx}m_{l_{sx}}}^{J_{a}M_{a}}\,C_{J_{s}M_{s}\,\,J_{x}M_{x}}^{s_{sx}m_{s_{sx}}}        
 \,{{\cal M}^{DWZR(prior)}},
\label{MDWBpreior1}
\end{align}
$l_{i\,j}$ is the orbital angular momentum of the bound state $(i\,j)$,  $\,\Gamma_{\tau (xA)}\,$  is the partial resonance width for the level $\tau$ in the channel $x+A$. 

In what follows, we assume that
$\,{\widehat {\bf{k}}_{xA}}\,$ is directed along the axis $\,z$, that is, $\,{Y_{{l_{xA}},{m_{{l_{xA}}}}}}\big({\widehat {\bf{k}}_{xA}}\big)= \sqrt{\frac{ {\hat l}_{xA}}{4\,\pi}}\,\delta_{m_{l_{xA}}\,0}$.

\begin{align}
& {\cal M}^{DWZR(prior)}= \int {d{{\rm {\bf r}}_{sx}}}\, \Psi_{ - {{\rm {\bf k}}_{sF}}}^{( + )}({{\rm {\bf r}}_{sx}})\,{\phi _{sx}}({{\rm {\bf r}}_{sx}})
\nonumber\\
&\times \Psi_{{\rm {\bf k_{aA}}}}^{( + )}(\frac{{{m_s}}}{{{m_a}}}{\rm {\bf r}}_{sx}).
\label{MDZRpr1}
\end{align}
is the DWBA amplitude, which does not depend on the resonant wave function of the resonance state $F_{\tau}^{*}$  and on the $V_{xA}$ potential.  $\, \Psi_{{\rm {\bf k_{aA}}}}^{( + )}(\frac{{{m_s}}}{{{m_a}}}{\rm {\bf r}}_{sx})\,$   and     
 $\, \Psi_{ - {{\rm {\bf k}}_{sF}}}^{( + )}({{\rm {\bf r}}_{sx}})\,$ are the distorted waves in the channels $a+A$ and $s+F$, respectively. The final-state distorted wave calculated at $\,k_{sF}= \sqrt{2\,\mu_{sF}\,E_{sF}}$. 
For simplicity, we assume that $l_{sx}=0,$  which usually is the case for the THM.
Equation (\ref{MDZRpr1}) looks like the zero-range DWBA (ZRDWBA).
The presence of the distorted waves in the initial and final states in the DWBA amplitude ${\cal M}^{DWZR(prior)}$
has a substantial impact on the angular distribution of $s$ and the absolute value of the transfer reaction amplitude.
We get the PWA amplitude if we replace the distorted waves with the plane waves.  

\begin{align}
&{\cal W}_{l_{xA}} = \Big[R_{ch}\,\frac{ {\partial {\ln[O_{{l_{xA}}}}({k_{xA}}{r_{xA}})]}}{{\partial {r_{xA}}}}
                                     \nonumber\\
&-1 - R_{ch}\,\frac{\partial{{\ln\,j_{{l_{xA}}}}({k_{xA}}{r_{xA}})}}{{\partial {r_{xA}}}}\Big]{\Big|_{{r_{xA}} = {R_{ch}}}}
\label{calWr1}
\end{align} 
is the off-shell factor, which  appears because the transferred particle $x$, see Fig.   \ref{fig_PWAdiagram},  is off-shell. 
$O_{{l_{xA}}}({k_{xA}}{R_{ch}})$  is outgoing scattered wave and $\,j_{{l_{xA}}}({k_{xA}}{R_{ch}})\,$ is the spherical Bessel function.

$W_{M_{B} M_{b}}^{\nu\,M_{F_{v}}}({\rm {\bf k}}_{bB}) $ is the vertex form factor for decay $F_{\nu} \to b+B$:
\begin{align}
&W_{ M_{B} M_{b}}^{\nu\,M_{F_{\nu}}}({\rm {\bf k}}_{bB}) = \sum\limits_{s_{bB}\, l_{bB}\,m_{s_{bB}}\,m_{l_{bB}} } \,C_{s_{bB}\,m_{s_{bB}} \,l_{bB}\,m_{l_{bB}}}^{ J_{F}\,M_{F_{\nu}}}\,
\nonumber\\
& \times  C_{J_{b}\,M_{b}\,\,J_{B}\,M_{B}}^{s_{bB}m_{s_{bB}}} 
\,Y_{l_{bB}\,m_{l_{bB}}}({\rm {\bf k}}_{bB})\,e^{i\,\delta_{l_{bB}}^{p}(k_{bB})}\,\sqrt{ \frac{ \mu_{bB}\,\Gamma_{\nu (bB)} }{k_{bB}}   }.
\label{vertexformfctor1}
\end{align}

Assembling Eqs. (\ref{MDWBACoul2}) - (\ref{vertexformfctor1})  we get 
\begin{align}
&\overline {\big|{\cal M}_R{\big|^2}} =\,\frac{{\hat J}_{F}}{{\widehat J}_{A}\,{\hat J}_{x}}\,\frac{{\hat l}_{xA}}{16\,\pi}\,
\,\frac{1}{\mu_{xA}}\,\frac{1}{k_{xA}}\,
\sum\limits_{ M} (C_{s\,M\,\,l_{xA}\,0}^{J_{F}\,M_{F}} )^{2}   \nonumber\\
&\times j_{l_{xA}}^{2}(k_{xA}\,R_{ch})\,{\cal W}_{l_{xA}}^{2}\,\big| {\cal M}^{DWZR(prior)} \big|^{2}  \nonumber\\
&\times \sum\limits_{M_{b}\,M_{B}}\,\Big|\,\sum\limits_{M_{F_{\nu}}}\, \sum\limits_{\nu\,\tau=1}^{N}\,W_{ M_{B} M_{b}}^{\nu\,M_{F_{\nu}}}({{\rm {\bf k}}_{bB}})\,[{\bf A}]_{\nu\,\tau}^{-1}\Gamma_{\tau\,(xA)}^{1/2} \Big|^{2}.
 \label{multilevelMR2}
 \end{align}

\subsubsection{Single-level, two-channel case}
\label{singlelevel1}

For the single-level case and $k_{sF}  \to k_{0(sF)}$
\begin{widetext}
\begin{align}
\overline {\big|{\cal M}_R{\big|^2}} 
 =& \frac{1}{ {\hat J}_{a} {\hat J}_{A} }\sum\limits_{ M_{F} M_{F}' M_{A} M_{a} M_{s} }\, M_{ M_{F} M_{s}; M_{A}M_{a}}({\rm {\bf { k}}}_{sF},{\rm {\bf k}}_{aA})\,[M_{ M_{F}' M_{s}; M_{A} M_{a} }({\rm {\bf { k}}}_{sF},{\rm {\bf k}}_{aA})]^*             \nonumber\\
& \times \frac{1}{  ( E_{0(sF)} - E_{sF} )^{2}  + {\Gamma ^2}/{4}  }\sum\limits_{ M_{B}M_{b} } W_{ M_{B} M_{b} }^{M_{F}} ({\rm {\bf k}}_{bB})\, \left[W_{ M_{B} M_{b} }^{M_{F}'}({\rm {\bf k}}_{bB}) \right]^*.
\label{THMreactampl1}
\end{align}
\end{widetext}
Here  $\,l_{bB}$ $ (m_{l_{bB}})$ is the $b-B$ relative orbital angular momentum
(its projection) in the resonance $F^{*}$,  $s_{bB}$ $(m_{bB})$ is the $b+B$ channel spin (its projection)  in the resonance state and $\delta_{l_{bB}}^{p}(k_{0(bB)})$ is the potential scattering phase shift in the $b+B$ channel, $k_{0(bB)} = \sqrt{2\,\mu_{bB}\,E_{0(bB)} }$.

\subsubsection{Coulomb factor}
\label{Coulombfactor}

Taking into account that
\begin{align}
\big|\Gamma[1+ i\,\eta] \big|^{2} = \frac{\pi\,\eta}{\sinh(\pi\,\eta)}
\label{Gammash1}
\end{align}
we get the Coulomb factor for narrow resonances \cite{mukkad2019}:
\begin{align}
|N_C|^{2}= & \frac{{\sinh[\pi ({\eta _{sb}} + {\eta _{sB}})]}}{{\sinh(\pi {\eta _{sb}})\sinh(\pi {\eta _{sB}})}}\,\frac{{\pi {\eta _{sb}}{\eta _{sB}}}}{{({\eta _{sb}} + {\eta _{sB}})}}\frac{{\pi {\eta _\zeta }}}{{\sinh(\pi {\eta _\zeta })}}                                           \nonumber\\
& \times |F( - i{\eta _{sB}}, - i{\eta _{sb}},1; - 1){|^2} \nonumber \\
& \times {\exp{\left[2\zeta \arctan \frac{{2({E_{0(bB)}} - {E_{bB}})}}{\Gamma }\right]}},
\label{NC21}
\end{align}
where
\begin{align}
\zeta=\eta _{sb} + \eta _{sB} - \eta _{{ 0}},
\label{zeta1}
\end{align}
$\eta_{j\,t}= (Z_{j}\,Z_{t}/137)\,\mu_{j\,t}/k_{j\,t}$, 
$\eta_{0}= Z_{s}\,Z_{F}\,\mu_{sF}/k_{0},$  $\,Z_{j}\,e$ is the charge of particle $j$. 
The presence of the factor $|N_{C}|^{2}$ shifts the resonance peak and the resonance shape line in the system $b+B$  due to the intermediate $s+F^{*}$ and the three-body ($s+b+B$) final-state Coulomb interactions.
The analytical expression for $N_{C}$ can be derived only for narrow resonances. The complications are caused by the final-state three-body Coulomb interactions \cite{mukkad2019}. 

It is convenient to integrate the triple DCS over $\,\Omega_{{\rm {\bf k}}_{bB}}\,$ to get the double DCS \cite{Dolinski}, which is expressed
in terms of the DCS of the reaction $\,a+ A \to s + F^{*}\,$ corresponding to the first step of the TH reaction.
However, in the case under consideration, due to the presence of the Coulomb factor $|N_C|^{2}$, the DCS obtained from integrating the triple DCS over $\,\Omega_{{\rm {\bf k}}_{bB}}\,$ cannot be expressed in terms of the DCS of the first step. The reason is that $N_C$ depends on the integration variable $\,\Omega_{{\rm {\bf k}}_{bB}}$.  However, in the following cases one can neglect this dependence: 
\begin{enumerate}
\item 
When $\,|\eta_{sb}| \ll  1\,$ and $\,\eta_{sB} \approx \eta_{{0}}.\,$ In this case, $\,|N_C| \approx 1\,$ and the integration over $\,\Omega_{{\rm {\bf k}}_{sF}}\,$ can be performed without any complications. \\
\item
 When $|\eta_{sb}| \ll 1$ and $m_{B} \gg m_{s},\,m_{b}$. Let us choose as independent variables the Galilean momenta
$\,{\rm {\bf k}}_{sF}\,$ and $\,{\rm {\bf k}}_{bB}$. Then one can write
\begin{align}
{\rm {\bf k}}_{sB}=  \frac{m_{B}\,M}{m_{sB}\,m_{bB}}\,{\rm {\bf k}}_{sF} + \frac{m_{s}}{m_{sB}}\,{\rm {\bf k}}_{bB}
\approx {\rm {\bf k}}_{sF}.
\label{ksB1}
\end{align}
Then $\eta_{sB}= (Z_{s}\,Z_{B}/137)\,\mu_{sB}/k_{sF}\,$ and $\,N_C\,$  does not depend on $\,{\rm {\bf k}}_{bB}\,$ and integration over
$\,\Omega_{{\rm {\bf k}}_{bB}}\,$  can be performed in a straightforward way.
\end{enumerate}

We assume that $|N_{C}|^{2} =1$.

\subsection{Double differential cross section}
\label{DoubleDCS1}

Taking into account Eq. (\ref{multilevelMR2}) and  integrating the triple DCS (\ref{TripleDCSspin1})  over $\,\Omega_{{\rm {\bf k}}_{bB}}\,$ using the orthogonality of the Clebsch-Gordan coefficients and  the spherical harmonics we get the double THM DCS
\begin{align}
&\frac{{{\rm d}^{2}\sigma^{THM} }}{{{\rm d}{\Omega _{{{\rm {\bf k}}_{sF}}}}{\rm d}{E_{sF}}}}  =
 \frac{{\hat J}_{F}}{{\widehat J}_{A}\,{\hat J}_{x}}
\,\frac{{\hat l}_{xA}}{16\,\pi}\,\frac{1}{(2\,\pi)^{3}}\,\frac{\mu_{aA}\,\mu_{sF}}{\mu_{xA}}\,
\frac{k_{sF}}{k_{aA}\,k_{xA}}\, \nonumber\\
&\times \sum\limits_{ M_F}\,(C_{s_{xA}\,M_{F}\,\,l_{xA}\,0}^{ J_{F}\,M_{F} } )^{2} \, j_{l_{xA}}^{2}(k_{xA}\,R_{ch})\,{\cal W}_{l_{xA}}^{2}\,\nonumber\\
&\times \big| {\cal M}^{DWZR(prior)} \big|^{2}  \,\Big| \sum\limits_{\nu\,\tau=1}^{N}\,\Gamma_{\nu (bB)}^{1/2}\, [{\bf A}]_{\nu\,\tau}^{-1}(E)\,\Gamma_{\tau\,(xA)}^{1/2} \Big|^{2}.
 \label{multlevDCS}
 \end{align}
Note that  Eq. (\ref{multlevDCS})   is  the THM double DCS proceeding through the intermediate resonance states.

For the single-level case the double THM DCS takes the form
\begin{align}
\frac{{{\rm d}^{2}\sigma^{THM} }}{{{\rm d}{\Omega _{{{\rm {\bf k}}_{sF}}}}{\rm d}{E_{sF}}}} = \frac{1}{{2\,\pi}}\frac{{{\Gamma _{bB}}}}{{{{({E_{0(bB)}} - {E_{bB}})}^2} + {{{\Gamma ^2}}}/{4}}}\frac{{d\sigma }}{{d{\Omega _{{{\rm {\bf k}}_{sF}}}}}},
\label{doubleDCS1}
\end{align}
where
\begin{align}
&\frac{{{\rm d}\sigma }}{{{\rm d}{\Omega _{{{\rm {\bf k}}_{sF}}}}}} =\frac{{{\mu _{aA}}{\mu _{sF}}}}{{4{\pi ^2}}}\frac{{{k_{sF}}}}{{{k_{aA}}}} \sum\limits_{M_{F}M_{s}M_{A}M_{a}}
\nonumber\\
& \times |{M_{ {M_F}{M_s};{M_A}{M_a}}}({\rm {\bf  k}}_{sF},{\rm {\bf k}}_{aA})|^2    \nonumber\\
& = \frac{\mu _{aA}\,\mu _{sF}}{(4\,\pi )^3}\,\frac{{\hat J}_{F}}{{\hat J}_{x}\,{\hat J}_{A}}\,\frac{{{k_{sF}}}}{{{k_{aA}}}}\,  
\frac{\Gamma_{xA}\,R_{ch} }{\mu _{xA}\,P_{l_{xA}}} \,j_{{l_{xA}}}^{2}({k_{xA}}{R_{ch}})               \nonumber\\
& \times {\cal W}_{l_{xA}}^{2}\, |{\cal M}^{DWZR(prior)}({\rm {\bf  k}}_{sF},\,{\rm {\bf k}}_{aA})|^{2}
\end{align}
is the DCS of the reaction $a+ A \to s+F^{*}$. 

In view of Eq. (\ref{GammaxAgamma1}) it is clear that the penetrability  factor  $P_{l_{xA}}$ is absent in the THM double DCS. 
It explains why the THM can be used to determine the astrophysical factor down to astrophysically relevant energies.

Integrating  the double DCS over $E_{sF}$ gives
\begin{align}
\int\limits_0^\infty\, {\rm d}{E_{sF}}\frac{{{\rm d}\sigma }}{{{\rm d}{\Omega _{{{\bf{k}}_{sF}}}}d{E_{sF}}}} = \frac{{{\Gamma _{bB}}}}{\Gamma }\frac{{d\sigma }}{{d{\Omega _{{{\bf{k}}_{sF}}}}}},
\end{align}
where $\Gamma_{bB}$ is the partial resonance width for the decay of the resonance to the channel $b+B$.

\subsection{THM astrophysical factor}
\label{THMSfactor1}

\subsubsection{$S$-factor for multi-level, two-channel case}

The astrophysical factor for the multilevel, two-channel case is (I remind that $E=E_{xA}$)
\begin{align}
&S(E)  = \frac{ {\hat J}_{F} }{ {\hat J}_{x}\,{\hat J}_{A}  }\,\frac{\pi}{2\,\mu_{xA}}\,e^{2\,\pi\,\eta_{xA}}      \nonumber\\
& \times  \lambda_{N}^{2}\,m_{u}^{2}\,\Big| \sum\limits_{\nu,\,\tau=1}^{N}\,\sqrt{\Gamma_{\nu (bB)}}\, [{\bf A}(E)]_{\nu\,\tau}^{-1}\,\sqrt{\Gamma_{\tau\,(xA)}} \Big|^{2}.                         
\label{Sfctrmultilvel1}
\end{align}
Here $\lambda_{N}$ is the nucleon Compton wave length, 
the reduced mass $\mu_{xA}$ is expressed in MeV, $\,m_{u}=931.5\,$ MeV  is the atomic mass.

Singling out this $S$-factor from THM double DCS we get
\begin{align}
&\frac{{{\rm d}^{2}\sigma }^{THM}}{{{\rm d}{\Omega _{{{\rm {\bf k}}_{sF}}}}{\rm d}{E_{sF}}}}  = e^{-2\,\pi\,\eta_{xA}} 
\,{\hat l}_{xA}\,\frac{\mu_{aA}\,\mu_{sF}}{2\,(2\,\pi)^{5}}\,\frac{k_{sF}}{k_{aA}}\,R_{ch}\,
\nonumber\\
&\times  P_{l_{xA}}^{-1}(E,R_{ch})\,  \sum\limits_{ M_F}\,(C_{s_{xA}\,M_{F}\,\,l_{xA}\,0}^{ J_{F}\,M_{F} } )^{2} \, j_{l_{xA}}^{2}(k_{xA}\,R_{ch})\,{\cal W}_{l_{xA}}^{2}\,\nonumber\\
&\times \lambda_{N}^{-2}\,m_{u}^{-2}\,
\big| {\cal M}^{DWZR(prior)} \big|^{2}\,S(E).
 \label{multileveldoubleDCS}
 \end{align}

\subsubsection{$S$-factor for the single-level, two channel case}

For the single-level, two channel case
\begin{align}
&S(E)  = \frac{ {\hat J}_{F} }{ {\hat J}_{x}\,{\hat J}_{A}  }\,\frac{\pi}{2\,\mu_{xA}}e^{2\,\pi\,\eta_{xA}}      
  \lambda_{N}^{2}\,m_{u}^{2}\,\frac{\Gamma_{bB}\,\Gamma_{xA}}{\big(E_{0(xA)}  - E \big)^{2}  + {\Gamma^{2}}/{4}}.
\label{SxA1}
\end{align}

This $S$-factor can be singled out from the THM double DCS. 
\begin{align}
&\frac{{{\rm d}^{2}\sigma }^{THM}}{{{\rm d}{\Omega _{{{\rm {\bf k}}_{sF}}}}d{E_{sF}}}} = S(E)\,e^{-2\,\pi\,\eta_{xA}}\,P_{l_{xA}}^{-1}(E,R_{ch})\,
\frac{{\hat l}_{xA}R_{ch} }{64\,\pi^{5}}\,
              \nonumber\\
& \times \lambda_{N}^{-2}\,m_{u}^{-2}\,\Big|{\cal W}_{l_{xA}}(E,R_{ch})\Big|^{2}\;
\frac{{{\rm d}\sigma }^{DWZR(prior)}}{{{\rm d}{\Omega _{{{\rm {\bf k}}_{sF}}}}}}
\label{doubleTHMDCS1}
\end{align}
and
\begin{align}
&\frac{{d\sigma }^{DWZR(prior)}}{{d{\Omega _{{k_{sF}}}}}} = \frac{{{\mu _{aA}}{\mu _{sF}}}}{{4{\pi ^2}}}\frac{{{k_{sF}}}}{{{k_{aA}}}} \nonumber\\
& \times \sum\limits_{M_{F}M_{s}M_{A}M_{a}}\Big|M_{{M_F}{M_s};{M_A}{M_a}}^{DWZR(prior)}({k_{0}}{{\bf{\hat k}}_{sF}},{{\bf{k}}_{aA}}){\Big|^2}
\label{singleDCS1}
\end{align}
is the DCS of the reaction $\,a+ A \to s+F^{*}\,$ populating the resonance state $\,F^{*}$.

The renormalization factor presented in Fig. \ref{fig_RenormSfctr1}   is the ratio of the DWBA differential cross
section given by Eq. (\ref{singleDCS1})  to the PWA differential cross section.

I presented a set of equations that can be used to analyze the THM reactions. Now we are in a position to discuss two THM reactions, which provide an indirect method to determine the astrophysical factor for 
the ${}^{12}{\rm C}+ {}^{12}{\rm C}$ fusion at the astrophysically relevant energies.

\section{ THM reactions induced by collision   $\boldsymbol{ {}^{14}{\rm N} + {}^{12}{\rm C}}$   }

The reactions 
\begin{align} 
{}^{14}{\rm N} + {}^{12}{\rm C} \to d+ {}^{24}{\rm Mg}^{*} \to \alpha_{0} (\alpha_{1}) + {}^{20}{\rm Ne} ({}^{20}{\rm Ne}^{*},  
\label{20Nealpha1}   \\
{}^{14}{\rm N} + {}^{12}{\rm C} \to d+ {}^{24}{\rm Mg}^{*} \to  p_{0}(p_{1}) + {}^{23}{\rm Na} ({}^{23}{\rm Na}^{*})
\label{23Nap1}
\end{align}
hade been used, see \cite{Tumino}, to determine the astrophysical factor for the carbon-carbon fusion at energies 
$E \geq 0.8$ MeV.  For this reaction $a={}^{14}{\rm N}\,\,s=d,\,\,x=A= {}^{12}{\rm C},\,\,F= {}^{24}{\rm Mg},\,b=p,\,\alpha,\,\,B= {}^{23}{\rm  Na},\,{}^{20}{\rm  Ne}.\,$  Also the orbital angular momentum  of the bound state ${}^{14}{\rm N}=(d\,{}^{12}{\rm C})$ is $l_{d\,{}^{12}{\rm C}}=0$. 

\subsection{$\boldsymbol{E_{{}^{14}{\rm N}} =30}$ MeV}
\label{E14N30MeV}

In \cite{Tumino}   the energy  of the ${}^{14}{\rm N}$ beam was $E_{{}^{14}{\rm N}}=30$ MeV corresponding to 
$E_{{}^{14}{\rm 14}\,{}^{12}{\rm C}} \approx 14$ MeV. Let us analyze the kinematic conditions of the THM reaction at this incident beam energy. \\
1.  $E_{{}^{14}{\rm N}\,{}^{12}{\rm C}} \approx 14$ MeV  is higher than the Coulomb barrier of $\approx 10$ MeV. This condition is necessary to avoid suppression of the THM DCS due to the Coulomb barrier in the entry channel ${}^{14}{\rm N}+ {}^{12}{\rm C}$. However, the relative energy $E_{d\,{}^{24}{\rm Mg}}$ remains well below the Coulomb barrier, especially at higher $E \equiv E_{{}^{12}{\rm C}\,{}^{12}{\rm C}}$. Large Coulomb parameter in the initial state of the THM reaction, $4.4$, and sub-Coulomb energies in the final state  makes the PWA used in \cite{Tumino}  invalid.\\
2. The second necessary condition of the THM is to utilize kinematics the most appropriate to measure the THM double DCS 
in the energy interval $0.8 \leq E \leq 2.66$ MeV (at energy $2.66$ MeV  the THM $S$-factor is normalized to the directly measured astrophysical fsctor) remaining close to the forward peak over the scattering angle $\theta_{sF}$ of the DCS $\frac{{{\rm d}\sigma }^{DWZR(prior)}}{{{\rm d}{\Omega _{{{\rm {\bf k}}_{sF}}}}}}$.  This peak corresponds to $p_{sx} < \kappa_{sx}$. To elaborate on the relationship between $p_{sx}$ and the THM double DCS, let us consider  a simple PWA, which can be obtained  by replacing in Eq. (\ref{MDZRpr1}) the distorted waves by the plane waves. Then  it is straightforward to see that the amplitude $ {\cal M}^{DWZR(prior)}$ is proportional to  $\,\phi_{sx}(p_{sx})$.  Hence the THM double DCS in the PWA  is proportional to $\phi_{sx}^{2}(p_{sx})$. Moreover, if the THM reaction is above the Coulomb barrier in the initial and final states, the $p_{sx}$ momentum distribution  extracted from the DWBA DCS should be close  to the momentum distribution given by  $\phi_{sx}^{2}(p_{sx})$ for $p_{sx}  <  \kappa_{sx}.$  

\begin{figure}
\includegraphics[height=17cm,width=0.5\textwidth]{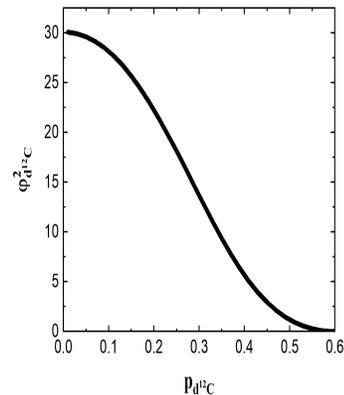}
\caption{Momentum distribution for the $(d\,{}^{12}{\rm C})$ bound state.}
\label{fig_14Nmomdistr}
\end{figure}

Fig. \ref{fig_14Nmomdistr}  shows the momentum distribution of $\phi_{d\,{}^{12}{\rm C}}^{2}(p_{d\,{}^{12}{\rm C}})$, which, as expected, has a peak at $p_{d\,{}^{12}{\rm C}}=0$. This peak is called a quasi-free  (QF) one  because  $p_{d\,{}^{12}{\rm C}}=0$ means that particles $s$ and $x$, being in the bound state ${}^{14}{\rm N}=(d\,{}^{12}{\rm C})$, are moving like  quasi-free  (non-interacting) particles with the relative zero velocity. The shape of the $s$-wave peak depends on $\,\kappa_{sx} = \sqrt{2\,\mu_{sx}\,\varepsilon_{sx}},\,$  $\,\varepsilon_{sx} = m_{s}+m_{x}-m_{a}$  is the binding energy of the bound state $a=(sx)$.  For the case under consideration the binding energy
 $\,\varepsilon_{d\,{}^{12}{\rm C}} =10.7\,$ MeV is, hence,   $\kappa_{d\,{}^{12}{\rm C}}= 1.7$ fm${}^{-1}$  are large and  the maximum 
of the momentum distribution of  $\phi_{d\,{}^{12}{\rm C}}^{2}(p_{d\,{}^{12}{\rm C}})$  occurs at $p_{d\,{}^{12}{\rm C}}<< \kappa_{d\,{}^{12}{\rm C}}$, that is, in the vicinity of $p_{d\,{}^{12}{\rm C}}=0$. For further analysis I assume that the QF kinematics  is also constrained by the condition $p_{d\,{}^{12}{\rm C}} << \kappa_{d\,{}^{12}{\rm C}}$. From the uncertainty principle follows that for $p_{sx} << \kappa_{sx}$ $\;r_{sx} > >  \kappa_{sx}^{-1}$, that is, a vicinity of the QF peak corresponds to larger distances between $d$ and ${}^{12}{\rm C}$, and we can  treat $d$ as a spectator with minimized impact on the ${}^{12}{\rm C}-{}^{12}{\rm C}$ interaction.

In Table \ref{table_EEdpsx1} are presented $E_{ d\,{}^{24}{\rm Mg} }$ and $p_{d\,{}^{12}{\rm C} }$ as
functions of the $E$ at $\theta=0$. We conclude that to cover the energy interval $0.8 \leq E \leq 2.7\, {\rm MeV}$ at fixed $E_{{}^{14}{\rm N}\,{}^{12}{\rm C}}$
one needs to vary $p_{d\,{}^{12}{\rm C}}$ in the interval $0 \leq p_{d\,{}^{12}{\rm C}} \leq 0.21\,$ fm${}^{-1}$. This interval is within the vicinity of the QF peak and at $p_{d\,{}^{12}{\rm C}} = 0.21\,$ the Fourier tansform $\,\phi_{d\,{}^{12}{\rm C}}^{2}(p_{d\,{}^{12}{\rm C}})$ drops only by a factor of $1.5$ from its peak value at $p_{d\,{}^{12}{\rm C}} =0$, see Fig. \ref{fig_14Nmomdistr}.
Another essential observation follows from comparing the columns $4$ and $5$.
A substantial difference off-shell momentum $p_{{}^{12}{\rm C}\,{}^{12}{\rm C}}\,$ 
and on-the-energy shell momentum $k_{ {}^{12}{\rm C}\,{}^{12}{\rm C}}= \sqrt{ 2\,\mu_{{}^{12}{\rm C}\,{}^{12}{\rm C}}\,E }$  underscores that off-the-energy shell effects 
 in the THM reaction (factor ${\cal W}_{l_{xA}}(E,R_{ch})$, see  Eq. (\ref{calWr1}) )  are very significant and on-the-energy shell approximation for the THM  double DCS is invalid.
\begin{table}
\caption{Kinematic energy-momentum calculations for $E_{ {}^{14}{\rm N} }= 30\,$ MeV.  $E_{ d\,{}^{24}{\rm Mg} }$ and  $p_{d\,{}^{12}{\rm C} }$ as
functions of the ${}^{12}{\rm C} -{}^{12}{\rm C}$ relative energy $E$. Equation (\ref{enconservTH1}) is used to calculate  $E_{d\,{}^{24}{\rm Mg}}$.   $p_{d\,{}^{12}{\rm C}}$ and 
$p_{{}^{12}{\rm C}\,{}^{12}{\rm C}}$ are calculated using Eq. (\ref{psxpxAEEaA1}) assuming that  $\theta=\arccos( {\rm{\bf {\hat k}}}_{d} \cdot {\rm {\bf {\hat k}}}_{{}^{14}{\rm N}})=0$.   $k_{ {}^{12}{\rm C}\,{}^{12}{\rm C}}$   is on-the-energy shell ${}^{12}{\rm C}- {}^{12}{\rm C}$ momentum.}
\label{table_EEdpsx2}
\begin{center}
\begin{tabular}{|c|c|c|c|c|}
\hline
 $E$ & $E_{d\,{}^{24}{\rm Mg}}\,$ &  $p_{d\,{}^{12}{\rm C}}\,$ & $p_{{}^{12}{\rm C}\,{}^{12}{\rm C}}\,$ &  $k_{ {}^{12}{\rm C}\,{}^{12}{\rm C}}\,$ \\ \hline
 (MeV)   & (MeV)  & (fm$^{-1}$)  & (fm$^{-1}$) & (fm$^{-1}$) \\ \hline
0.8 & 2.93 & 0.211537   & 1.82631 & 0.479292 \\ \hline
0.9 & 2.83 & 0.202781   & 1.83069 & 0.508366 \\ \hline
1.  & 2.73 & 0.193868   & 1.83515 & 0.535864 \\ \hline
1.1 & 2.63 & 0.18479    & 1.83969 & 0.562019 \\ \hline
1.2 & 2.53 & 0.175539   & 1.84431 & 0.58701  \\ \hline
1.3 & 2.43 & 0.166102   & 1.84903 & 0.61098  \\ \hline
1.4 & 2.33 & 0.156469   & 1.85385 & 0.634043 \\ \hline
1.5 & 2.23 & 0.146627   & 1.85877 & 0.656297 \\ \hline
1.6 & 2.13 & 0.136562   & 1.86381 & 0.677821 \\ \hline
1.7 & 2.03 & 0.126258   & 1.86896 & 0.698682 \\ \hline
1.8 & 1.93 & 0.115696   & 1.87424 & 0.718938 \\ \hline
1.9 & 1.83 & 0.104858   & 1.87966 & 0.738638 \\ \hline
2.  & 1.73 & 0.0937186  & 1.88523 & 0.757827 \\ \hline
2.1 & 1.63 & 0.0822526  & 1.89096 & 0.776541 \\ \hline
2.2 & 1.53 & 0.0704291  & 1.89688 & 0.794815 \\ \hline
2.3 & 1.43 & 0.0582125  & 1.90298 & 0.812679 \\ \hline
2.4 & 1.33 & 0.0455608  & 1.90931 & 0.830158 \\ \hline
2.5 & 1.23 & 0.0324238  & 1.91588 & 0.847276 \\ \hline
2.6 & 1.13 & 0.0187409  & 1.92272 & 0.864056 \\ \hline
2.7 & 1.03 & 0.00443808 & 1.92987 & 0.880515 \\ \hline
 \hline
\end{tabular}
\label{table_EEdpsx1}
\end{center}
\end{table}

\begin{figure}
\includegraphics[width=3.0in,height=3.5in]{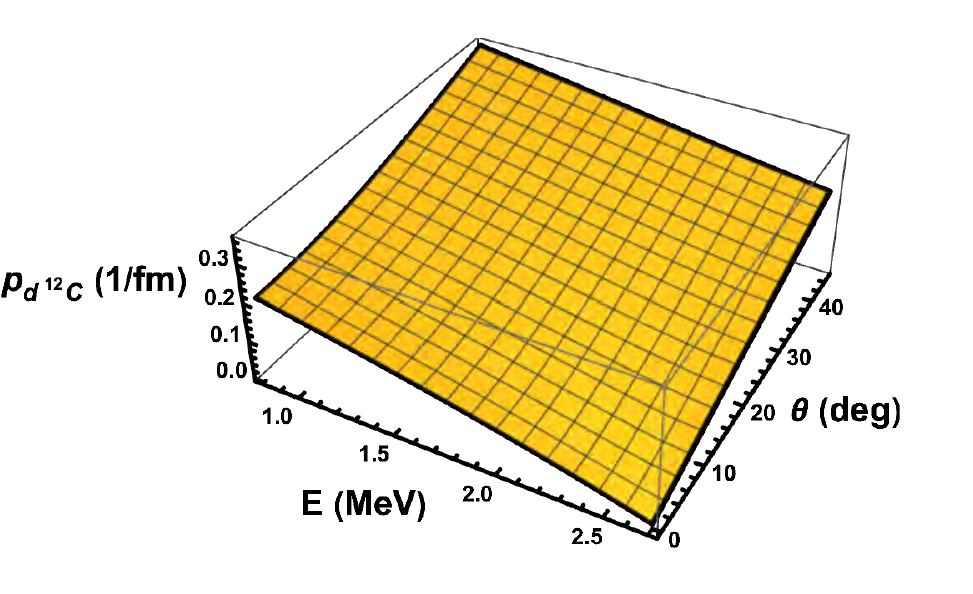}
  \caption{$p_{p_{d\,{}^{12}{\rm C}}}$ as a function of $E$ and $\theta$ for $E_{{}^{14}{\rm N}} = 30\,$ MeV. }
\label{fig_3Dplot}
\end{figure}

$3$D plot in Fig. \ref{fig_3Dplot} shows the momentum $p_{p_{d\,{}^{12}{\rm C}}}$ as a function of $E$ and $\theta$. 
 It allows one to select  optimal kinematic conditions at each energy $E$: smaller angles $\theta$ (the angle  between ${}{\rm {\bf k}}_{{}^{14}{\rm N}}$ and  ${\rm {\bf k}}_{d}$) provide smaller $p_{d\,{}^{12}{\rm C}}$, that is, higher THM double DCS.  At $p_{d\,{}^{12}{\rm C}}=0.3\,$ fm$^{-1}$  $\,\phi_{d\,{}^{12}{\rm C}}^{2}(p_{d\,{}^{12}{\rm C}})$ drops  by about a factor of two compared to its peak value.  

\begin{figure}
\includegraphics[width=3.0in,height=3.5in]{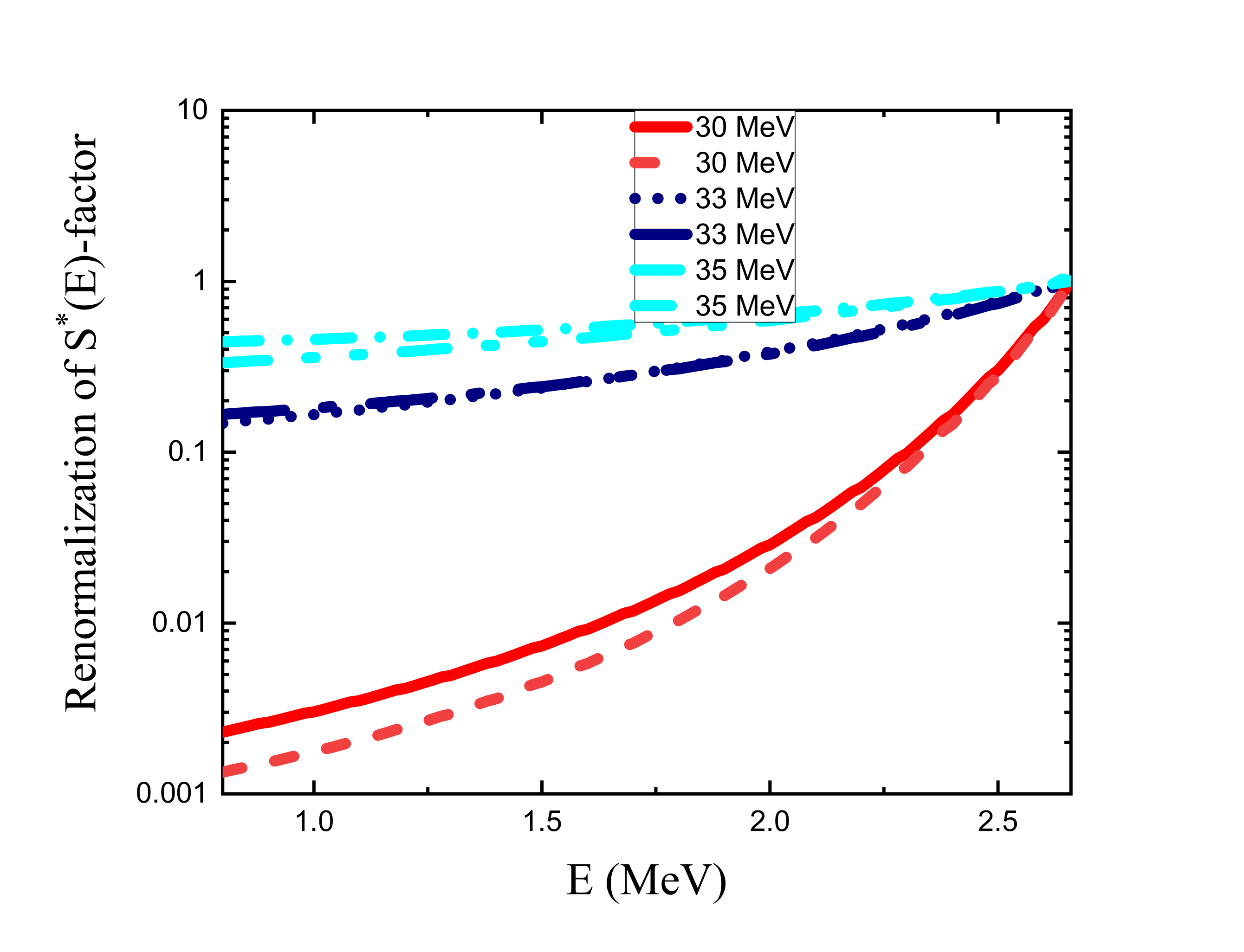}
  \caption{Renormalization factor of the generalized THM astrophysical factor caused by Coulomb-nuclear distortions for three different 
  $E_{{}^{14}{\rm N}}$ energies: $30,\,33\,$ and $35\,$ MeV. There are two lines for each energy: the low line is the renormalization factor caused  by the pure Coulomb distortions and the upper line is caused by the Coulomb-nuclear distortions. }
\label{fig_RenormSfctr1}
\end{figure}

Thus a comprehensive analysis show that the beam energy $\,E_{{}^{14}{\rm N}}= 30\,$ MeV could provide an optimal kinematics to cover the astrophysically relevant energies $ 0.8 \leq E \leq 2$ MeV for low $\,p_{d\,{}^{12}{\rm C}}$. However, a very strong effect of the Coulomb-nuclear distortions neglected in \cite{Tumino} completely changes both the absolute value of DCS, see Fig. \ref{fig_RenormSfctr1}, and the deuteron angular distributions \cite{muk2019,muk2020}. These distortions is the main obstacle 
to use the THM reaction at $\,E_{{}^{14}{\rm N}}=30\,$ MeV.  
Fig. \ref{fig_RenormSfctr1} shows the renormalization factors of the modified THM astrophysical factors at different beam energies. We see that at $E_{{}^{14}{\rm N}} =30$ MeV at low $E$ the renormalization factor is about $0.001$. However, at the beam energy of $35$ MeV
at low $E$, the renormalization factor is about two orders of magnitude larger than at $30$ MeV.  

\subsection{$\boldsymbol{E_{{}^{14}{\rm N}} = 35}$ MeV}
\label{E14N35MeV}

Since the $30\,$ MeV  beam is too low, I will repeat all the previous calculations at $E_{{}^{14}{\rm N}}=35\,$ MeV.
\begin{table}
\caption{Kinematic energy-momentum calculations for $E_{ {}^{14}{\rm N} }= 35\,$ MeV. Notations are the same as in Table \ref{table_EEdpsx1}. }
\label{table_EEdpsx2}
\begin{center}
\begin{tabular}{|c|c|c|c|c|}
\hline
 $E$ & $E_{d\,{}^{24}{\rm Mg}}\,$ &  $p_{d\,{}^{12}{\rm C}}\,$ & $p_{{}^{12}{\rm C}\,{}^{12}{\rm C}}\,$ &  $k_{ {}^{12}{\rm C}\,{}^{12}{\rm C}}\,$ \\ \hline
 (MeV)   & (MeV)  & (fm$^{-1}$)  & (fm$^{-1}$) & (fm$^{-1}$) \\ \hline
0.8 & 5.08 & 0.352861 & 1.90051 & 0.479292 \\ \hline
0.9 & 4.98 & 0.346213 & 1.90381 & 0.508366 \\ \hline
1.  & 4.88 & 0.339497 & 1.90714 & 0.535864 \\ \hline
1.1 & 4.78 & 0.332712 & 1.9105  & 0.562019 \\ \hline
1.2 & 4.68 & 0.325857 & 1.9139  & 0.58701  \\ \hline
1.3 & 4.58 & 0.318927 & 1.91734 & 0.61098  \\ \hline
1.4 & 4.48 & 0.311921 & 1.92082 & 0.634043 \\ \hline
1.5 & 4.38 & 0.304837 & 1.92433 & 0.656297 \\ \hline
1.6 & 4.28 & 0.297671 & 1.92789 & 0.677821 \\ \hline
1.7 & 4.18 & 0.290422 & 1.93148 & 0.698682 \\ \hline
1.8 & 4.08 & 0.283084 & 1.93512 & 0.718938 \\ \hline
1.9 & 3.98 & 0.275657 & 1.93881 & 0.738638 \\ \hline
2.  & 3.88 & 0.268135 & 1.94254 & 0.757827 \\ \hline
2.1 & 3.78 & 0.260516 & 1.94632 & 0.776541 \\ \hline
2.2 & 3.68 & 0.252796 & 1.95015 & 0.794815 \\ \hline
2.3 & 3.58 & 0.24497  & 1.95404 & 0.812679 \\ \hline
2.4 & 3.48 & 0.237033 & 1.95798 & 0.830158 \\ \hline
2.5 & 3.38 & 0.228982 & 1.96197 & 0.847276 \\ \hline
2.6 & 3.28 & 0.220811 & 1.96603 & 0.864056 \\ \hline
2.7 & 3.18 & 0.212514 & 1.97014 & 0.880515 \\ \hline
\end{tabular}
\label{table_EEdpsx2}
\end{center}
\end{table}
The second column of Table \ref{table_EEdpsx2}  shows that at $\,E_{ {}^{14}{\rm N} }= 35\,$ MeV  all $E_{d\,{}^{24}{\rm 24}}$ are higher than the Coulomb barrier (only on the higher end of $E\;$ energy $E_{d\,{}^{24}{\rm 24}}$ approaches the Coulomb barrier of $3\,$ MeV). Taking into account that the initial energy $E_{{}^{14}{\rm N}\,{}^{12}{\rm C}} = 16.15\,$ MeV 
is higher than the initial Coulomb barrier $10\,$ MeV we conclude that  the angular distribution of the deuterons is forward peaked. Hence the experimental momentum distribution of the deuterons extracted from the THM double DCS for all $E$ and 
$p_{d\,{}^{12}{\rm C}}$ from  Table \ref{table_EEdpsx2}  should be similar to the momentum distribution  in Fig. \ref{fig_14Nmomdistr}. 

\begin{figure}
\includegraphics[width=3.0in,height=3.5in]{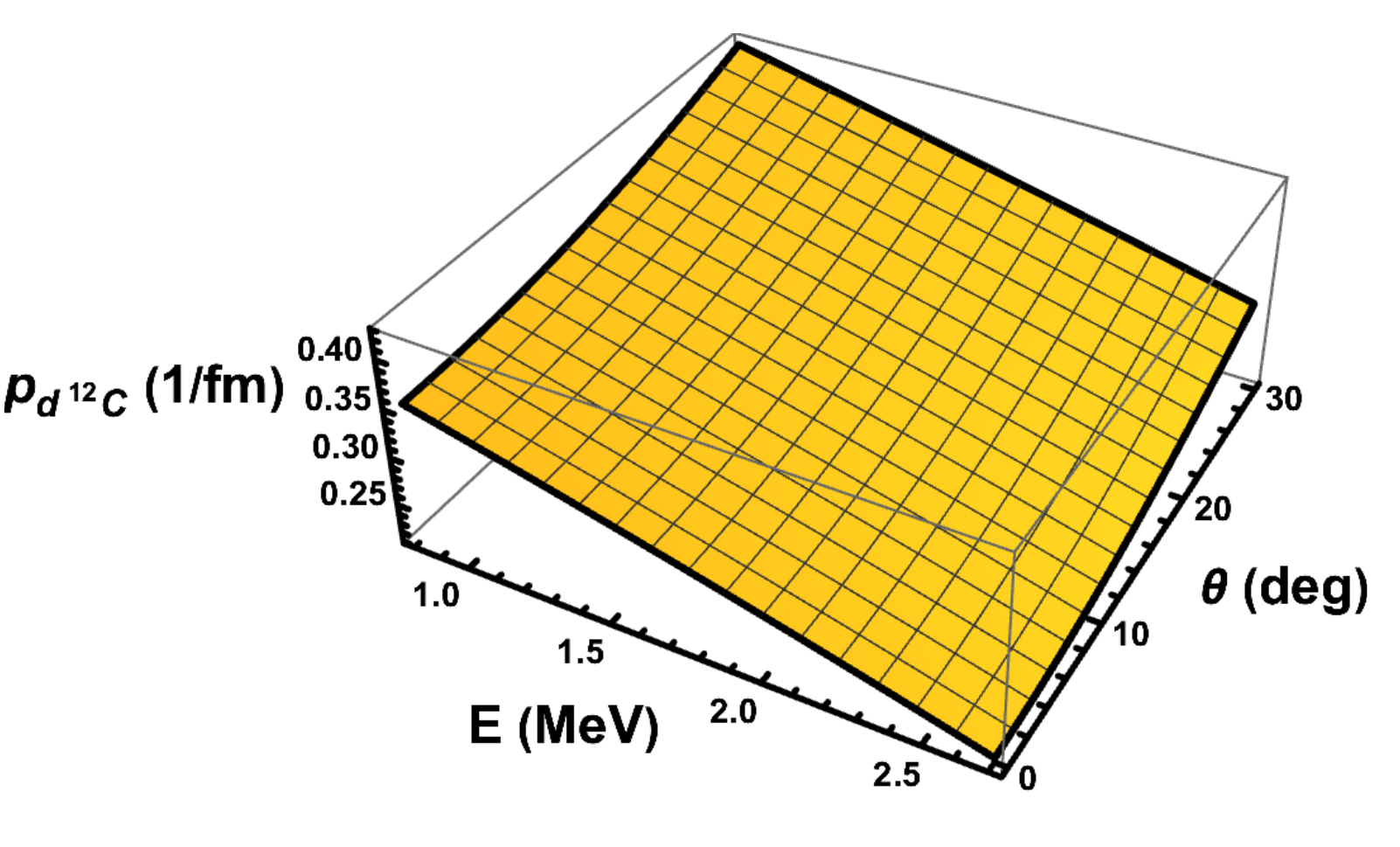}
  \caption{$p_{p_{d\,{}^{12}{\rm C}}}$ as a function of $E$ and $\theta$ for $E_{{}^{14}{\rm N}} = 35\,$ MeV.}
\label{fig_3Dplot35MeV}
\end{figure}

$3$D plot in Fig. \ref{fig_3Dplot35MeV}  shows the momentum $p_{d\,{}^{12}{\rm C}}$ as a function of $E$ and $\theta$ for $E_{{}^{14}{\rm N}}=35\,$ MeV.

Let us select two important resonance energies: $E=0.9\,$ and $1.5\,$ MeV. 
From Tables \ref{table_EEdpsx1}  and \ref{table_EEdpsx2} follows that for these resonances:  for $\,E_{{}^{14}{\rm N}}= 30\,$ MeV
$p_{d\,{}^{12}{\rm C}}=0.20$ and $0.15$ fm${}^{-1}$, respectively;  for 
$\,E_{ {}^{14}{\rm N} }= 35\,$ MeV  $\,p_{d\,{}^{12}{\rm C}}=0.35\,$  and $0.30$ fm${}^{-1}$, respectively. 
From Fig. \ref{fig_14Nmomdistr}  we find that   $ \phi_{d\,{}^{12}{\rm C}}^{2}(0.35\,{\rm fm}^{-1})/
\phi_{d\,{}^{12}{\rm C}}^{2}(0.2\,{\rm fm}^{-1})=0.43$ and $ \phi_{d\,{}^{12}{\rm C}}^{2}(0.30\,{\rm fm}^{-1})/
\phi_{d\,{}^{12}{\rm C}}^{2}(0.15\,{\rm fm}^{-1})=0.53$, respectively. 
 These ratios show the decrease of the THM double DCS at two critical resonance energies when the energy of the incident beam of ${}^{14}{\rm N}$ increases from $30$ MeV to $35$ MeV.   Meantime,  from Fig. \ref{fig_RenormSfctr1} follows that the ratio of the renormalization factors $R_{35\,{\rm MeV}}(0.9\, {\rm MeV})R_{30\,{\rm MeV}}(0.9\, {\rm MeB}) = 333.3$ and 
$R_{35\,{\rm MeV}}(1.5\, {\rm MeV})R_{30\,{\rm MeV}}(1.5\, {\rm MeB}) = 115.6$. 
Thus the drop of  $\phi_{d\,{}^{12}{\rm C}}^{2}(p_{d\,{}^{12}{\rm C}})$  with increase of the energy $\,E_{{}^{14}{\rm N}}\,$  is well compensated by
the increase of the renormalization factor: the total gain for the energy $\,E_{{}^{14}{\rm N}}\,$ increase from $30\,$ to $35\,$ MeV is
$\,110\,$  for $\,E=0.9\,$ MeV  and  $\,61\,$ for $\,E=1.5\,$ MeV. 
Thus an increase of the beam energy from $\,30\,$ MeV \cite{Tumino} to $\,35\,$ MeV will increase the THM double DCS because the outgoing deuterons become above the Coulomb barrier.  Besides, the angular distribution of the deuterons will be forward peaked and the THM experiment can be repeated 
at  $\,E_{{}^{14}{\rm N}}=35\,$ MeV  avoiding the  problems appeared for
$E_{{}^{14}{\rm N}}=30\,$ MeV.

\section{ THM reaction induced by collision   $\boldsymbol{ {}^{13}{\rm C} + {}^{12}{\rm C}}$   }
\label{THM13C12C1}

Now I consider another THM reactions suitable for determination of the $S$-factors  for the carbon-carbon fusion: 
\begin{align} 
{}^{13}{\rm C} + {}^{12}{\rm C} \to n+ {}^{24}{\rm Mg}^{*} \to \alpha_{0} (\alpha_{1}) + {}^{20}{\rm Ne} ({}^{20}{\rm Ne}^{*}),  
\label{20Nealpha1}   \\
{}^{13}{\rm C} + {}^{12}{\rm C} \to n+ {}^{24}{\rm Mg}^{*} \to  p_{0}(p_{1}) + {}^{23}{\rm Na} ({}^{23}{\rm Na}^{*}).
\label{23Nap1}
\end{align}
In the  previous section notations we need to  replace $a$ and $s$ with $a={}^{13}{\rm C}$  and $s=n$; also now the orbital angular momentum of the bound state ${}^{13}{\rm C}= (n\,{}^{12}{\rm C})\,$  $\,l_{n\,{}^{12}{\rm C}}=1$ and the bound-state wave number
$\kappa_{n\,{}^{12}{\rm C}} = 0.47$. 
Thus these will be the first THM reactions in which the Trojan horse particle is the $\,p\,$-wave bound state. 
Since the spectator in reactions (\ref{20Nealpha1}) and (\ref{23Nap1})  is neutron,  there are no Coulomb interactions in the intermediate and final state of the THM reactions, which suppressed the THM double DCS in the previous section with the deuteron as a spectator. 

\begin{figure}
\includegraphics[width=2.5in,height=3.0in]{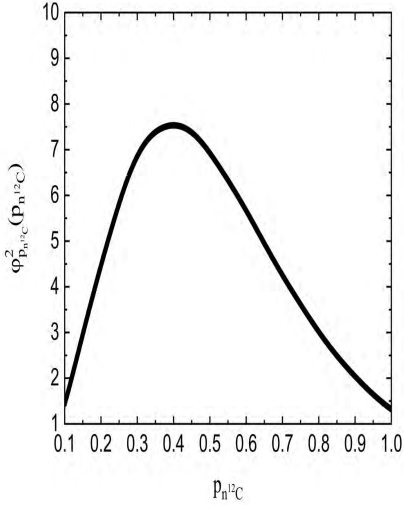}
\caption{Momentum distribution for the $(n\,{}^{12}{\rm C})$ bound state.}
\label{fig_n12Cmomentumdistr}
\end{figure}

The momentum distribution of $\,\phi_{n\,{}^{12}{\rm C}}^{2}(p_{n\,{}^{12}{\rm C}})\,$ shown in Fig. \ref{fig_n12Cmomentumdistr},  due to $\,l_{n\,{}^{12}{\rm C}}=1,\,$ is peaked at $\,p_{n\,{}^{12}{\rm C}}=0.4\,$ fm${}^{-1}\,$ rather then at $\,p_{d\,{}^{12}{\rm C}}=0\,$ for the $s$-wave bound state. 

In Table \ref{table_EEdpsx3} are presented $E_{ n\,{}^{24}{\rm Mg} }$ and $p_{n\,{}^{12}{\rm C} }$ as
functions of  $E$ at $\theta=0$ at $E_{{}^{13}{\rm C}} =29\,$ MeV. We conclude that to cover the energy interval $0.8 \leq E \leq 2.7\, {\rm MeV}$ at fixed $E_{{}^{13}{\rm C}}=29\,$ MeV  one needs to vary $p_{n\,{}^{12}{\rm C}}$ in the interval $0.38 \leq p_{n\,{}^{12}{\rm C}} \leq 0.46\,$ fm${}^{-1}$. This interval  is lower than $\kappa_{n\,{}^{12}{\rm C}}= 0.47$ fm${}^{-1}.\,$ Within this interval the Fourier transform $\,\phi_{d\,{}^{12}{\rm C}}^{2}(p_{n\,{}^{12}{\rm C}})$   changes very little, see Fig. \ref{fig_n12Cmomentumdistr}.

\begin{figure}
\includegraphics[width=3.0in,height=3.5in]{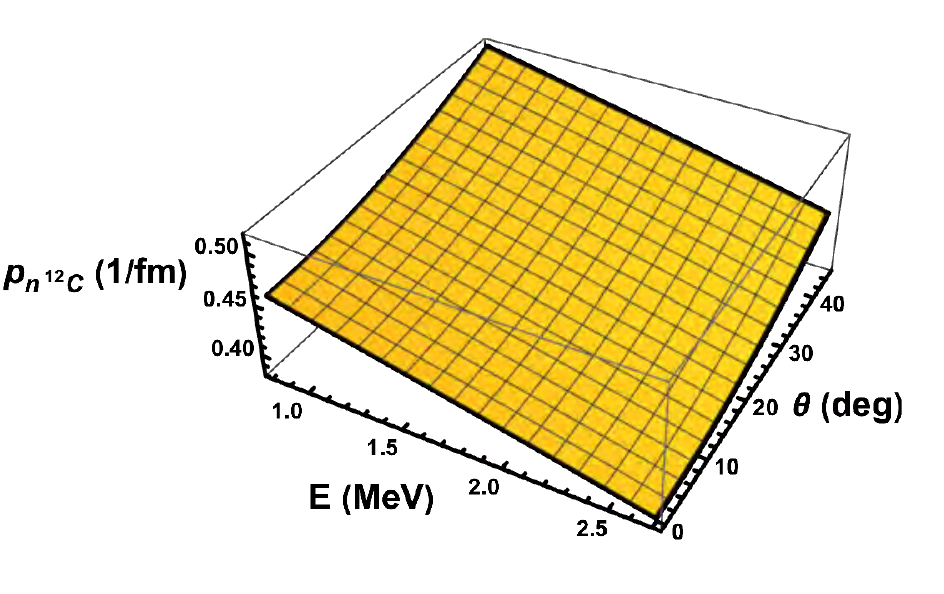}
  \caption{$p_{n\,{}^{12}{\rm C}}$ as a function of $E$ and $\theta$ for $E_{{}^{13}{\rm C}} = 29\,$ MeV.}
\label{fig_3Dplot29MeV}
\end{figure}

Also, as it was the case for ${}^{14}{\rm N} + {}^{12}{\rm C}$,  from comparing the columns $4$ and $5$ we see that the off-shell effects are very significant.
\begin{table}
\caption{Kinematic energy-momentum calculations for $E_{ {}^{13}{\rm C} }= 29\,$ MeV.  $E_{ n\,{}^{24}{\rm Mg} }$ and  $p_{n\,{}^{12}{\rm C} }$ as
functions of the ${}^{12}{\rm C} -{}^{12}{\rm C}$ relative energy $E$. Equation (\ref{enconservTH1}) is used to calculate  $E_{n\,{}^{24}{\rm Mg}}$.   $p_{n\,{}^{12}{\rm C}}$ and 
$p_{{}^{12}{\rm C}\,{}^{12}{\rm C}}$ are calculated using Eq. (\ref{psxpxAEEaA1}) assuming that  $\theta=\arccos( {\rm{\bf {\hat k}}}_{n} \cdot {\rm {\bf {\hat k}}}_{{}^{13}{\rm C}})=0$.   $k_{ {}^{12}{\rm C}\,{}^{12}{\rm C}}$   is on-the-energy shell ${}^{12}{\rm C}- {}^{12}{\rm C}$ momentum.}
\begin{tabular}{|c|c|c|c|c|}
\hline
 $E$ & $E_{n\,{}^{24}{\rm Mg}}\,$ &  $p_{n\,{}^{12}{\rm C}}\,$ & $p_{{}^{12}{\rm C}\,{}^{12}{\rm C}}\,$  &  $k_{ {}^{12}{\rm C}\,{}^{12}{\rm C}}\,$   \\ \hline
 (MeV)   & (MeV)  & (fm$^{-1}$)  & (fm$^{-1}$) & (fm$^{-1}$) \\ \hline
0.8 & 8.15 & 0.457724 & 1.73267 & 0.479292 \\ \hline
0.9 & 8.05 & 0.453942 & 1.73454 & 0.508366 \\ \hline
1.  & 7.95 & 0.450137 & 1.73642 & 0.535864 \\ \hline
1.1 & 7.85 & 0.446308 & 1.73832 & 0.562019 \\ \hline
1.2 & 7.75 & 0.442455 & 1.74022 & 0.58701  \\ \hline
1.3 & 7.65 & 0.438577 & 1.74214 & 0.61098  \\ \hline
1.4 & 7.55 & 0.434673 & 1.74407 & 0.634043 \\ \hline
1.5 & 7.45 & 0.430743 & 1.74602 & 0.656297 \\ \hline
1.6 & 7.35 & 0.426787 & 1.74797 & 0.677821 \\ \hline
1.7 & 7.25 & 0.422804 & 1.74994 & 0.698682 \\ \hline
1.8 & 7.15 & 0.418794 & 1.75193 & 0.718938 \\ \hline
1.9 & 7.05 & 0.414755 & 1.75393 & 0.738638 \\ \hline
2.  & 6.95 & 0.410687 & 1.75594 & 0.757827 \\ \hline
2.1 & 6.85 & 0.40659  & 1.75797 & 0.776541 \\ \hline
2.2 & 6.75 & 0.402463 & 1.76001 & 0.794815 \\ \hline
2.3 & 6.65 & 0.398305 & 1.76207 & 0.812679 \\ \hline
2.4 & 6.55 & 0.394116 & 1.76414 & 0.830158 \\ \hline
2.5 & 6.45 & 0.389895 & 1.76623 & 0.847276 \\ \hline
2.6 & 6.35 & 0.385641 & 1.76833 & 0.864056 \\ \hline
2.7 & 6.25 & 0.381354 & 1.77045 & 0.880515 \\ \hline
\end{tabular}
\label{table_EEdpsx3}
\end{table}
$3$D plot  in Fig. \ref{fig_3Dplot29MeV}  shows that for small neutron scattering angles $\,p_{n\,{}^{12}{\rm C}} <\kappa_{n\,{}^{12}{\rm C}}.\,$

\section{Summary}
\label{Summary1}
The update of the current status of the modified astrophysical $S^{*}(E)$-factors for the carbon-carbon fusion is given. The latest two direct measurements in \cite{STELLA} and \cite{Tan} show disagreement at energies $E \geq 2.1\,$ MeV. It is  not feasible in the near future that  direct measurements can reach the Gamow window  $1.2 \leq E \leq 1.8$ MeV. Nowadays, the only way to reach these energies  is to use the indirect THM. 
In the THM only the energy dependence of the astrophysical factor is extracted. To get their absolute value one needs to normalize the THM data to direct ones at higher energies where reliable direct measurements are available.  For the carbon-carbon fusion the normalization energy was $E \approx 2.6\,$ MeV.  Thus the THM measurements should cover quite a broad energy interval, $0.8 \leq E \leq  2.7\,$ MeV. Reconciliation of direct measurements at
$E=2.1\,$ MeV would help THM experiments: the normalization to direct measurements can be performed at $2.1\,$ MeV. 

Some important THM equations are presented, which are needed to apply the THM. To determine  the $S^{*}$-factors for the carbon-carbon fusion for the astrophysically relevant energy interval  $0.8 \leq E \leq 2.0\,$ MeV   different THM reactions have been analyzed. Among them are two THM reactions induced by the collision $\,{}^{14}{\rm N} + {}^{12}{\rm C}\,$ at $E_{{}^{14}{\rm N}} =30\,$ MeV \cite{Tumino}. and $E_{{}^{14}{\rm N}} =35\,$ MeV.  It is shown that  the higher energy beam 
allows one to avoid difficulties in the THM  experiment  \cite{Tumino} at $E_{{}^{14}{\rm N}} =30\,$ MeV and can be used to 
extract the $S^{*}$-factors for the carbon-carbon fusion.

Also,  the kinematics of the THM reaction induced by the ${}^{13}{\rm C} + {}^{12}{\rm C}$ collision is analyzed and is shown that this reaction can be used to determine the $S^{*}$-factors for the carbon-carbon fusion. 

\acknowledgments{A. M. M. acknowledges a support from the U.S. DOE Grant No. DE-FG02-93ER40773, and the NNSA Grant No. DENA000384.}

\begin{center}

\end{center}

\end{document}